\documentclass[english,prl, twocolumn]{revtex4}
\usepackage{amsfonts}
\usepackage{amsmath}
\usepackage{graphicx}
\usepackage{color}
\usepackage{babel}
\usepackage{hyperref}
\usepackage{epsfig,bm}
\usepackage{graphicx}
\usepackage{amsmath,amssymb,psfrag,textcomp}
\usepackage{amsthm,upgreek}

\newcommand{\be}{\begin{equation}}
\newcommand{\ee}{\end{equation}}
\newcommand{\bd}{\begin{displaymath}}
\newcommand{\ed}{\end{displaymath}}

\newcommand{\beeq}[1] {\begin{equation}\begin{split}#1\end{split}\end{equation}}

\begin{document}
\title{A weighted message-passing algorithm to estimate volume-related properties of random polytopes}
\author{Francesc Font-Clos}
\affiliation{Centre de Recerca Matem\`atica, Edifici C, Campus UAB, E-08193 Bellaterra (Barcelona), Spain}
\author{Francesco Alessandro Massucci}
\author{Isaac P\'erez Castillo}
\affiliation{Department of Mathematics, King's College London, WC2R 2LS United Kingdom}
\begin{abstract}
In this letter, we introduce a novel message-passing algorithm for a class of problems which can be mathematically understood as estimating  volume-related properties of random polytopes. Unlike the usual approach consisting in approximating the real-valued cavity marginal distributions by a few parameters, we propose a weighted message-passing algorithm to deal with the entire function.  Various alternatives of how to implement our approach are discussed and numerical results for random polytopes are compared with results using the Hit-and-Run algorithm.
\end{abstract}
\maketitle
As it is well-known, old and new interesting problems such as Gardner's optimal capacity problem for continuous synaptic couplings \cite{Gardner1988}, von Neumann's problem of linear economics \cite{VN}, estimation of reaction fluxes from mass-balance equations \cite{Kauffman2003}, or reconstruction techniques in compressed sensing \cite{Krzakala2011}, can be related, in one way or another, to estimating volume-related properties of random polytopes.  In the so-called H-representation, a polytope is defined as the set of points $\bm{x}=(x_1,\ldots,x_N)\in D\subseteq\mathbb{R}^N$  encapsulated by $M$ hyperplanes $\{(\bm{\xi}^\mu,\gamma^\mu)\}_{\mu=1}^M$, with $\bm{\xi}^\mu=(\xi^\mu_1,\ldots,\xi^\mu_N)$ the normal vector of the $\mu$-hyperplane. From all possible questions related to polytopes in H-representation,  we are particularly interested in that of the volume $V$ and its projection onto each axis relative to $V$ or, in order words, the marginal pdf $P_i(x_i)$. These two quantities can  mathematically be written as:
\begin{eqnarray}
V& =& \int_{D} d\bm{x}\prod_{\mu=1}^M \Theta \left(\sum_{i=1}^N \xi_i^\mu x_i - \gamma^\mu\right)\label{partition}\,,\\
P_i(x_i)&=&\frac{1}{V}\int_{D} d\bm{x}_{\setminus i}\prod_{\mu=1}^M \Theta \left(\sum_{j=1}^N \xi_j^\mu x_j - \gamma^\mu\right)\,,
\end{eqnarray}
where $\Theta(x)$ is the Heaviside step function and $\bm{x}_{\setminus i}$ denotes the vector $\bm{x}$ without component $i$.

Finding efficient methods to obtain reliable estimates for the marginals $P_i(x_i)$ has been, and still is, the main objective of past and present research. In the context of metabolic networks, the area we will focus on to present our work, these methods can roughly be divided into either theoretical approaches of some sort, or Monte Carlo simulations. In the first class of methods, we find Flux Balance Analysis (FBA) \cite{Kauffman2003}, which consists in approximating the whole volume $V$ of plausible solutions of the mass-balance equations by a single point, which is selected by optimising an objective function \footnote{In FBA, one deals with set of equalities instead of inequalities, but the approach is the same. For introduction of FBA see \cite{Kauffman2003}.}. FBA provides reliable results when applied to elementary micro-organisms in simple milieu conditions (see, for instance, \cite{Segre2002,Kauffman2003}), but the collapsing of $V$ to a single point is too drastic for more complex micro-organisms. To explore the whole volume $V$ one usually relies on Monte Carlo simulations, as the one introduced in \cite{Price2004}, or the Hit-and-Run algorithm \cite{Berbee1987}. These two algorithms provide a uniform sampling of $V$, but they are restricted to small networks \footnote{The mixing-time of the Hit-and-Run algorithm goes like $\mathcal{O}(N^3)$, which makes it impractical for large networks}. For larger networks the MinOver$^{+}$ algorithm has shown to be  very promising \cite{DeMartino2007,DeMartino2009,DeMartino2010b}, with the sampling becoming more uniform the larger the network is.

Here we want to re-explore the possibility of using message-passing algorithms to estimate $P_i(x_i)$. Message-passing equations, that is cavity equations, were already derived in \cite{DeMartino2007} for the von Neumann problem, and in \cite{Braunstein2008} for mass-balance problem. As the dynamical variables are continuous, it was thought that having a message-passing algorithm for the exact marginals was a daunting task \cite{DeMartino2007}, even though fast Fourier transform has been nicely applied in some cases \cite{Braunstein2008}. 

Let us see how it is possible to introduce a weighted message-passing algorithm for the entire cavity density marginals. Having  applications in the area of metabolism in mind, we explain our methodology in the context of the  von Neumann model applied to metabolic networks \cite{VN, Martelli2009, DeMartino2010b}. The application to similar problems is straightforward. In here, a metabolic network is a system of $i=1,\ldots,N$ coupled chemical reactions which produces and consumes $\mu=1,\ldots,M$ metabolites and exchanges them with the environment.  Following von Neumann's work \cite{VN,DeMartino2007}, as some of the  the input will be used to produce output, the ratio of input produced to output consumed cannot be larger than a global growth rate $\rho$, that is
\beeq{
\sum_{i=1}^N \left(b_i^\mu - \rho a_i^\mu \right)x_i \geq \gamma^\mu\,,\quad\quad \mu=1,\ldots,M\,,
\label{VonNeumann}
}
where $A=(a_i^\mu)$ and $B=(b_i^\mu)$  are the input and output matrices (e.g. stoichiometric coefficients), $x_i$ is the rate of chemical reaction $i$, and  $\bm{\gamma}=(\gamma^1,\ldots,\gamma^M)$ is the vector of exchanged rates (associated to e.g. transport of metabolites). Given a metabolic network, that is, given a set of stoichiometric coefficients and exchanged rates, we want to find the volume of solutions $V$ to the set of inequalities \eqref{VonNeumann} at the optimal growth rate. As we do not want to get distracted by unnecessary complicacies, we denote $\xi^\mu_i\equiv \xi^\mu_i(\rho)=b_i^\mu - \rho a_i^\mu$, ignoring the dependence on $\rho$ and calling $\bm{\xi}=(\xi_i^\mu)$ the stoichiometric matrix.\\
In practical situations, the stoichiometric matrix is generally diluted and it is not unreasonable to consider the problem of the volume for diluted random polytopes, meaning that the corresponding bipartite graph associated to the stoichiometric matrix $\bm{\xi}$ is tree-like for both $i$-nodes and $\mu$-nodes. The volume of solutions $V$ is given by \eqref{partition}, with $D$ being the domain region of the reaction rates. This domain is usually determined by the biochemistry of each reaction, that is, reaction rates are usually bounded $x_i\in[x_i^{\text{min}},x_i^{\text{max}}]$. For sake of simplicity we drop the domain from the notation. Using the cavity method, we obtain the following set of coupled cavity equations \cite{DeMartino2007,Font2011}:
\begin{widetext}
\begin{eqnarray}
m_\mu^{(i)}(s_i) &=& \frac{1}{m_\mu^{(i)}}\int  d \bm{s}_{\partial \mu \setminus i} ~\delta\left(\gamma^\mu-\eta_i^\mu s_i-\sum_{j\in \partial \mu\setminus i}\eta_j^\mu s_j\right) \prod_{j\in \partial \mu\setminus i}  P^{(\mu)}_j(s_j)\,,\quad \mu=1,\ldots,M\,,\quad i\in\partial\mu\label{cavityequations2.1}\\
 P^{(\mu)}_i(s_i)& =&\frac{1}{Z_i^{(\mu)}}\prod_{\nu \in \partial i\setminus \mu} m_\nu^{(i)}(s_i)\,,\quad i=1,\ldots,N\,,\quad \mu\in \partial i
\label{cavityequations2.2}
\end{eqnarray}
\end{widetext}
where $m_\mu^{(i)}$ and $Z_i^{(\mu)}$ are normalisation constants and the notation e.g. $j\in \partial \mu$ means considering all reactions which participate in the production or consumption of metabolite $\mu$. Here $\bm{s}=(x_1,\ldots,x_N,z_1,\ldots,z_M)$, where the $z-$variables have been introduced to write the step functions as integrated Dirac deltas, while the matrix $\bm{\eta}$ corresponds to the augmented stoichiometric matrix resulting from this transformation. Although this is not a necessary mathematical step, it will be useful for the subsequent discussion.\\
The actual marginals $P_i(x_i)$ are given by
\beeq{
P_i(x_i) &=\frac{1}{Z_i}\prod_{\nu \in \partial i} m_\nu^{(i)}(x_i)\,,\quad i=1,\ldots,N\,,
}  
while the meaning of  $m_\mu^{(i)}(x_i)$ in eq. \eqref{cavityequations2.1} is the following: assuming that the marginals $ \{P^{(\mu)}_{\ell}(x_\ell)\}_{\ell\in\partial \mu\backslash i}$ when reaction $i$ has been removed  are known,  the probability that, by fixing the value at node  $i$ to $x_i$ the inequality $\mu$ will be fulfilled, is precisely $m_\mu^{(i)}(x_i)$.\\
At this point, one would be tempted to use the Dirac delta in eq. \eqref{cavityequations2.1} in an iteration process with assignments $(\gamma^\mu-\sum_{j\in \partial \mu\setminus i}\eta_j^\mu s_j)/\eta_i^\mu\to s_i$ as it is usually done in the standard disordered-averaged cavity equations \cite{Mezard2001}. There are two differences, though: (i) the set of equations \eqref{cavityequations2.1} is still on an instance; (ii) due to the integration domain, not all possible assignments $(\gamma^\mu-\sum_{j\in \partial \mu\setminus i}\eta_j^\mu s_j)/\eta_i^\mu\to s_i$ are allowed. Fortunately, the acceptance region in eq. \eqref{cavityequations2.1} for $m_\mu^{(i)}(s_i)$ is easy to determine as it corresponds to restricting the domain $D$  by the hyperplane $\mu$. Within the acceptance region, we can rewrite equation \eqref{cavityequations2.1} as follows:
\begin{widetext}
\beeq{
m_\mu^{(i)}(s_i) &=\frac{1}{m_\mu^{(i)}} \left[\prod_{j=1}^{k_\mu^{(i)}}\int_{\mathcal{R}^{(\mu)}_j(s_{\ell_1},\ldots,s_{\ell_{j-1}};\gamma^\mu)}  d s_{\ell_{j}} P^{(\mu)}_{\ell_j}(s_{\ell_j}|s_{\ell_1},\ldots,s_{\ell_{j-1}};\gamma^\mu)\right]\\
&\times ~\delta\left(\gamma^\mu-\eta_i^\mu s_i-\sum_{j=1}^{k_\mu^{(i)}}\eta_j^\mu s_{\ell_j}\right) \prod_{j=1}^{k_\mu^{(i)}}\omega_{j}^{(\mu)}(s_{\ell_1},\ldots,s_{\ell_{j-1}};\gamma^\mu)
\label{cavityequations3}
}
with
\beeq{
P^{(\mu)}_{\ell_j}(s_{\ell_j}|s_{\ell_1},\ldots,s_{\ell_{j-1}};\gamma^\mu)=\frac{P^{(\mu)}_{\ell_j}(s_{\ell_j})}{\omega_{j}^{(\mu)}(s_{\ell_1},\ldots,s_{\ell_{j-1}};\gamma^\mu)}\,,\quad \omega_{j}^{(\mu)}(s_{\ell_1},\ldots,s_{\ell_{j-1}};\gamma^\mu)=\int_{\mathcal{R}^{(\mu)}_j(s_{\ell_1},\ldots,s_{\ell_{j-1}};\gamma^\mu)}ds_{\ell_j}P^{(\mu)}_{\ell_j}(s_{\ell_j})
}
\end{widetext}
Here $k_\mu^{(i)}=|\partial \mu \setminus i|$ and $\ell_{j}\in \partial \mu \setminus i$ for $j=1,\ldots,k_\mu^{(i)}$. The new expression \eqref{cavityequations3} invites us to propose the following updating method to solve numerically this equation:
\begin{enumerate}
\item Draw $s_{\ell_1}$ with probability $P^{(\mu)}_{\ell_1}(s_{\ell_1}|\gamma^\mu)$, draw $s_{\ell_2}$ with probability $P^{(\mu)}_{\ell_2}(s_{\ell_2}|s_{\ell_1};\gamma^\mu)$, $\ldots$, draw $s_{\ell_{k_\mu^{(i)}}}$ with probability $P^{(\mu)}_{\ell_{k_\mu^{(i)}}}(s_{\ell_{k_\mu^{(i)}}}|s_{\ell_1},\ldots,s_{\ell_{k_\mu^{(i)}-1}};\gamma^\mu)$.
\item Assign $s_i\leftarrow(\gamma^\mu-\sum_{j\in \partial \mu\setminus i}\eta_j^\mu s_j)/\eta_i^\mu$ with weight $\Omega^{(\mu)}_i\equiv\prod_{j=1}^{k_\mu^{(i)}}\omega_{j}^{(\mu)}(s_{\ell_1},\ldots,s_{\ell_{j-1}};\gamma^\mu)$
\end{enumerate}
With this weighted iteration in mind, it is possible to solve numerically the set of equations \eqref{cavityequations2.1} and  \eqref{cavityequations2.2} directly. We propose two methods: (i) method of histograms. Here one represents $\{P^{(\mu)}_{\ell_j}(s_{\ell_j})\}$ with histograms. Firstly, we use the weighted iteration method in equation \eqref{cavityequations2.1} to obtain a population of size $\mathcal{N}$ of pairs $\{(s_{i,\alpha},\Omega^{(\mu)}_{i,\alpha})\}_{\alpha=1}^{\mathcal{N}}$ from which to construct a histogram for $m_\mu^{(i)}(s_i)=(1/\sum_{\alpha=1}^{\mathcal{N}}\Omega^{(\mu)}_{i,\alpha})\sum_{\alpha=1}^{\mathcal{N}}\Omega^{(\mu)}_{i,\alpha}\delta(s_i-s_{i,\alpha})$. Then use equation \eqref{cavityequations2.2} to recalculate the histograms of $\{P^{(\mu)}_{\ell_j}(s_{\ell_j})\}$ from the histograms of $\{m_\mu^{(i)}(s_i)\}$; (ii) method of the weighted populations. It is possible to circumvent the step of using and constructing histograms during the iteration process. One simply works directly with a set of equations for the cavity marginals $\{P^{(\mu)}_{\ell_j}(s_{\ell_j})\}$, by combining eqs. \eqref{cavityequations2.1} and \eqref{cavityequations2.2} . Each of these functions is represented by a weighted population $\{(s_{j,\alpha},\Gamma^{(\mu)}_{j,\alpha})\}_{\alpha=1}^{\mathcal{N}}$ of size $\mathcal{N}$. Note that in this case, one Dirac delta appearing in the combined expression gives the assignment for $s_j$ while at the same time, locks the other Dirac deltas, providing an extra weight.

To assess  the validity of our approach we have run some numerical checks and compared them with Monte Carlo simulations using the Hit-and-Run algorithm. We entirely focus our numerics in the methods of histograms and postpone a more detailed analysis on the method of weighted populations for another occasion \cite{Font2011}.  We have created a random metabolic network of $N=25$ reactions and $M=10$ metabolites. The in-  and  out-degree  distributions for reaction and metabolite nodes are chosen to be Poisson with average degree 1.5 and 3.75, respectively.  The stoichiometric coefficients $\xi_i^\mu$ as well as the exchange fluxes $\gamma^\mu$ are chosen randomly, but ensuring that the volume of solutions is not empty. Finally, we assume that the domain $D=[0,1]^{25}$.

\begin{figure}[h]
\begin{center}
\includegraphics[width=1.0\columnwidth, height=9cm]{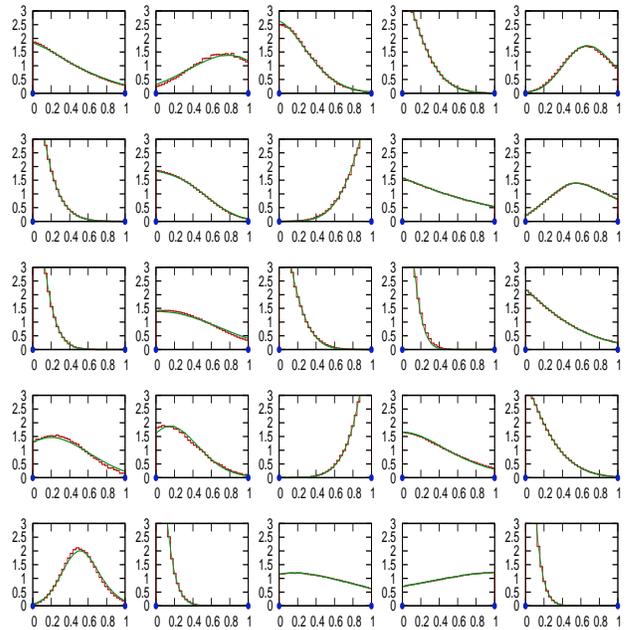}
\end{center}
\caption{Result for the marginals $P_i(x_i)$ for $i=1,\ldots,25$ (shown in red and ordered from left to right and up to bottom) using the weighted iteration algorithm within the histogram method for a random metabolic network of $N=25$ reactions and $M=10$ metabolites. Our numerical results are compared with Monte Carlo simulations using the Hit-and-Run algorithm (green line). Finally, the solid blue  circles are the results of the belief-propagation algorithm for the boundaries of the marginals $\{P_i(x_i)\}$.}
\label{fig1}
\end{figure}
Results are summarised in fig.~\ref{fig1}, which shows all marginals $\{P_{i}(x_i)\}_{i=1}^{25}$ obtained by the weighted message-passing algorithm, together with the Monte Carlo simulations using the Hit-and-Run algorithm. The comparison is fairly excellent, considering that the cavity equations are based on the Bethe approximation, which does not generally hold for a random graph of finite size, particularly this small.

\begin{figure}[h]
\begin{picture}(175,175)
\put(75,0){$x_i/b$}
\put(-45,85){$f_{i}(x)$}
\put(-35,170){\includegraphics[height=.95\columnwidth, angle=-90]{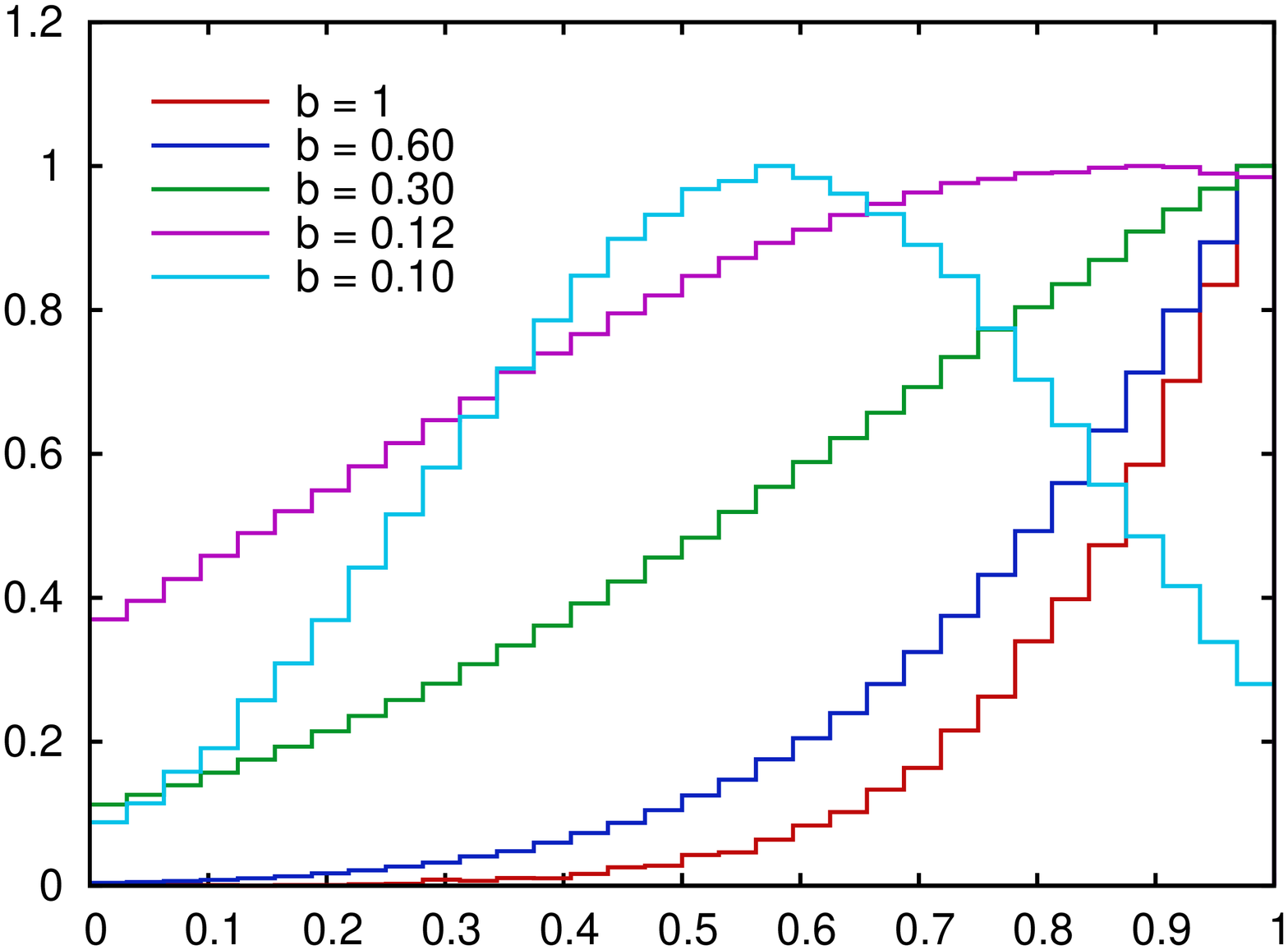}}
\end{picture}
\caption{Results for the marginal $P_{i}(x_i)$ for $i=18$ and by varying the hypercube $[0,b]^{25}$ for $b=1.0,0.60,0.30,0.12$, and $b=0.10$. For a visually easier comparison, the  horizontal and vertical axes have been rescaled as $x_i/b$ and $f_i(x)=P_{i}(x_i)/M_i$ with $M_i=\text{max}_{x \in[0,b]}P_{i}(x)$, respectively.}
\label{fig2}
\end{figure}

Looking at the results in fig. \ref{fig1} we also note that, although originally the domain for each reaction rate is the interval $[0,1]$, some reaction rates seem to have a smaller domain region. This can be due because either the polytope further  constrains the original domain or rather because there exists a positive but very small probability. It is important to distinguish between what cannot happen and what is unlikely to happen, particularly in metabolism, where a perturbed network can be brought into a region of originally unlikely events (i.e. a disease). To shed some light into this matter, we notice that the set of equations  \eqref{cavityequations2.1} and \eqref{cavityequations2.2} suggests a way to write a belief-propagation algorithm for the domains $[a_i,b_i]\ni x_i$ of  the marginals $P_i(x_i)$. (see \cite{Font2011} for details). The results are also reported in fig. \ref{fig1} as solid blue circles and  indicate that, for this particular instance, the probability in some regions for some variables is very small. To explore further into this matter and to access these regions, we have performed a very simple perturbation of the original metabolic network consisting in changing the domain $D$ by  $[0,b]^{25}$ for $b=1.0,0.60,0.30,0.12$, and $b=0.10$.  Results are shown in fig. \ref{fig2} for variable 18, where we can see that not only the probability increases, but also that the shape of the pdf changes as well in a non-trivial way.

\begin{figure}[h]
\begin{picture}(150,150)
\put(75,-5){$a$}
\put(-40,75){$b$}
\put(-40,0){\includegraphics[width=.95\columnwidth ]{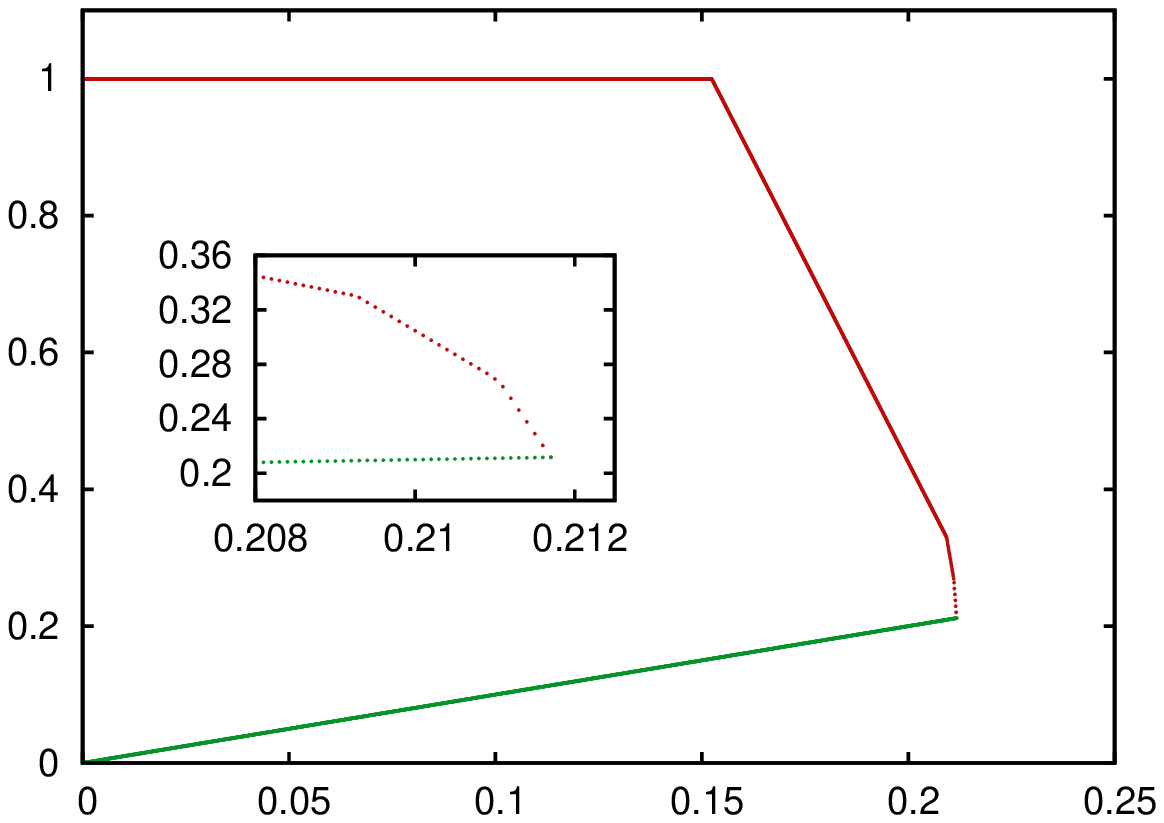}}
\end{picture}
\caption{Behaviour of domains $[a,b]\ni x_i$ of marginal $P_i(x_i)$ variable $i=22$ as the end point $a$ is varied for a perturbed domain $[a,1]^{25}$. The inset captures a more detailed behaviour of the region  where the end points meet.}
\label{fig3}
\end{figure}

It is interesting to keep in mind that since a marginal $P_i(x_i)$ corresponds to projecting the volume of the polytope onto the $x_i$-axis, restrictions in the polytope may affect both boundaries $[a,b]\ni x_i$. To follow track of these changes the message-passing equations for the domain \cite{Font2011} are most convenient. As an example, in fig. \ref{fig3} we have plotted how the domain $[a,b]$ of the marginal $P_i(x_i)$ for variable $22$ changes as $a$ in the original domain $[a,1]^{25}$ of the polytope is varied.\\

In this letter we have presented a weighted message-passing algorithm to deal with a class of problems which consists in counting solutions of a set of equations (either equalities or inequalities). The method is designed to deal with the exact cavity marginals, rather than approximating these with a set of parameters. We have introduced two ways of implementing the method and discussed here one of them, the method of histograms. This method has admittedly the drawback of having to approximate the functions with histograms, but avoids the rejection region characteristic of the problem.

Among the various research lines were are currently considering, we would like to mention two. Firstly, we are currently exploring the implementation of the algorithm by the method of weighted populations \cite{Font2011}. Like in the standard replica symmetric equations in the ensemble, the cavity marginals are naturally represented by a population of random variables. In this way, we avoid having to introduce any discretization of the marginals. Secondly, we are extending this approach to the more interesting case of the double von Neumann problem \cite{DeMartino2011}, which is able to incorporate information on the energy balance of the chemical reactions.

\bibliographystyle{apsrev}
\bibliography{bibliography}
\end{document}